\documentclass[a4paper]{article}
\usepackage[english]{babel}
\usepackage[utf8x]{inputenc}
\usepackage[T1]{fontenc}
\usepackage{CJK}
\usepackage[a4paper,top=3cm,bottom=2cm,left=3cm,right=3cm,marginparwidth=1.75cm]{geometry}
\usepackage{amsmath}
\usepackage{graphicx}
\usepackage[colorinlistoftodos]{todonotes}
\usepackage[colorlinks=true, allcolors=blue]{hyperref}
\usepackage{cite}
\usepackage{indentfirst}
\usepackage{authblk}
\newcommand{\bean}{\begin{eqnarray}}
\newcommand{\eean}{\end{eqnarray}}
\newcommand{\bea}{\begin{eqnarray*}}
\newcommand{\eea}{\end{eqnarray*}}
\newcommand{\eqs}[1]{Eqs. (\ref{#1})}
\newcommand{\eq}[1]{Eq. (\ref{#1})}
\newcommand{\meq}[1]{(\ref{#1})}
\newcommand{\fig}[1]{Fig. \ref{#1}}
\newcommand{\ppa}[2]{\left(\frac{\partial}{\partial #1}\right)^{#2}}
\newcommand{\pp}[2]{\frac{\partial #1}{\partial #2}}
\newcommand{\eqn}{&=&}

\begin{document}
\begin{CJK*}{UTF8}{gbsn}
\title{Critical phenomena in gravitational collapse of Husain-Martinez-Nunez  scalar field}
\author[1]{Xiaobao Wang\thanks{bao@mail.bnu.edu.cn}}
\author[2]{Xiaoning Wu\thanks{wuxn@amss.ac.cn}}
\author[3]{Sijie Gao\thanks{Corresponding author: sijie@bnu.edu.cn}}
\affil[1]{Department of Physics, Beijing Normal University, Beijing , China,100875}
\affil[2]{Institute of Applied Mathematics, Academy of Mathematics and System Science, Chinese Academy of Sciences, P.O. Box 2734, Beijing, China, 100080}

\maketitle
\CJKindent

\abstract
We construct analytical models to study the critical phenomena in gravitational collapse of the Husain-Martinez-Nunez massless scalar field. We first use the  cut-and-paste technique to  match the conformally flat solution ($c=0$ ) onto an outgoing Vaidya solution. To guarantee the continuity of the metric and the extrinsic curvature, we prove that the two solutions must be joined at a null hypersurface and the metric function in Vaidya spacetime must satisfy some constraints. We find that the mass of the black hole in the resulting spacetime takes the form $M\propto (p-p^*)^\gamma$, where the critical exponent $\gamma$ is equal to $0.5$. For the case $c\neq 0$, we show that the scalar field must be joined onto two pieces of Vaidya spacetimes to avoid a naked singularity. We also derive the power-law mass formula with $\gamma=0.5$. Compared with previous analytical models constructed from a different scalar field with continuous self-similarity, we obtain the same value of $\gamma$. However, we show that the solution with $c\neq 0$ is not self-similar. Therefore, we provide a rare example that a scalar field without self-similarity also possesses the features of critical collapse.

\section{ Introduction}

Gravitational collapse is the main reason of various galactic structures and it remains one of the most interesting and fundamental problems in general relativity. The end state of gravitational collapse could be a black hole, naked singularity or flat spacetime. In a seminal work by Choptuik  \cite{bibitem7}, some intriguing and universal properties concerning the formation of black holes from massless scalar fields were found. This is called the critical phenomenon. Particularly, near the threshold, the black hole mass can always be expressed in the form of power-law:
\bean
M \propto (p -p^*)^\gamma \,, \label{sform}
\eean
for $p>p^*$, where  $p$ is a parameter of the initial data to the threshold of black hole formation.
Numerical simulations have shown that that the critical exponent $\gamma$ is equal to $0.5$ for  solutions with continuous self-similarity (CSS) and $\gamma \approx 0.37$ for solutions with discrete  self-similarity (DSS). Details about the critical phenomenon  can be learned in  \cite{bibitem3a,bibitem3}.

In addition to numerical calculation, analytical models were also built to explore the critical phenomena.
Patrick R. Brady\cite{bibitem6}  studied an exact one parameter family of scalar field solutions which exhibit critical behaviours when black hole forms.
J. Soda and K. Hirata\cite{bibitem6a}  analytically studied the collapse of continuous self-similar scalar field in higher dimensional spacetimes and found a general formula for the critical exponents which agrees with the exponent $\gamma=0.5$ for $n=4$.  A. Wang et al \cite{bibitem5}  constructed an analytical model by  pasting the BONT model (a massless scalar field) with the Vaidya model. They demonstrated that the black hole mass obeys the power law with  $\gamma=0.5$.
 A. Wang et al \cite{bibitem2} also  analytically studied the gravitational collapse of a massless scalar field with conformal flatness.  They showed that the mass of the  black hole without self-similarity turns on with finite nonzero values. Recent developments regarding the critical phenomenon can be found in Refs. \cite{rct1,rct2,rct3,rct4,rct5,rct6,rct7}.

In this paper, we investigate the critical phenomena associated with an exact scalar field solution discovered by Husain-Martinez-Nunez (HMN) \cite{bibitem1}.  The HMN solution is interesting because it represents a black hole in a FLRW universe\cite{bibitem52,bibitem53,bibitem56}. It was pointed out that the HMN solution brings in new phenomenology(S-curve) of apparent horizon\cite{bibitem57}. Moreover, the conformally transformed HMN spacetime can be a inhomogeneous vacuum solution in Brans-Dick theory \cite{bibitem48,bibitem44,bibitem45}. Following the treatment in \cite{bibitem5}, we match the HMN solution onto an outgoing Vaidya solution along a null hypersurface. This is to guarantee that the black hole mass is finite. Usually, the hypersurface connecting the two parts of the spacetime is a thin shell, i.e., the extrinsic curvature across the hypersurface is discontinuous. By applying the
Darmois-Israel formula \cite{bibitem21}, one can find the relationship between the jump of the extrinsic curvature and the surface stress-energy tensor of the thin shell. However, by properly choosing the function $m(U)$ in the Vaidya metric, the extrinsic metric can be continuous across the null surface. Therefore, no thin shell forms and the resulting spacetime can be at least $C^1$.

After the matching, we calculate the apparent horizon and define the black hole mass as the Komar mass at the intersection of the apparent horizon and the null hypersurface. We first study the case $c=0$ where the HMN solution is conformally flat and has CSS. The mass of the black hole is found in the power-law form
$M\propto  \sqrt{-a}$ for $a<0$, which means $\gamma=0.5$. When $a=0$, the black hole disappear and the spacetime becomes Minkowski.

 The case of $c\neq 0$ is more complicated. This solution has no self-similarity. We still find $M\propto  \sqrt{-a}$ for $a<0$. Differing from the case of $c=0$, we show that the limiting spacetime ($a=0$) possesses a naked singularity with non-zero ADM mass. Therefore, there is a mass gap between the black hole spacetime ($a<0$) and the spacetime with naked singularity $a=0$.

The paper is organized as follows.
In section \ref{HMN},  we briefly introduce the Husain-Martinez-Nunez (HMN) scalar field solution. In section \ref{cfcc}, by using the cut-and-paste method, we match the conformally flat HMN solution ($c=0$ with CSS) onto an outgoing Vaidya spacetime at both timelike and null hypersurfaces to construct an analytical model. It turns out that only the matching along the null hypersurface can guarantee the continuity of the metric and extrinsic curvature across the surface. Then, we use this analytical model to  study the critical phenomenon  and derive the  mass formula.  In section \ref{gscc}, we join a general HMN ($a\neq0,c\neq0$) with two outgoing Vaidya spacetimes. We show that the mass of the black hole approaches zero for $a<0$. We also find that the critical spacetime ($a=0$) possesses a naked singularity with nonzero ADM mass. Concluding remarks are given in Section \ref{sec-con}.
In Appendix \ref{appendix}, we prove that the HMN solution  has CSS only when and $a\neq 0$ and $c=0$.

\section{ Husain-Martinez-Nunez (HMN) spacetime}\label{HMN}
The Husain-Martinez-Nunez spacetime \cite{bibitem1} satisfies the Einstein-scalar field equations
\begin{eqnarray}\label{m1}
G_{ab}&=&8\pi T_{ab}\,, \\
T_{ab}&=&\nabla_a\Phi\nabla_b\Phi-\frac{1}{2}g_{ab} g^{cd}\nabla_c\Phi\nabla_d\Phi\,.
\end{eqnarray}
The spherically symmetric solution is given by  \footnote{ Without loss of generality, we have set $b=1$ in the original metric in \cite{bibitem1}. }
\begin{eqnarray}\label{m2}
ds^2&=&(a t+1)\bigg{[}-\left(1-\frac{2c}{r}\right)^{\alpha} dt^2+\left(1-\frac{2c}{r}\right)^{-\alpha}dr^2
+r^2\left(1-\frac{2c}{r}\right)^{1-\alpha}d\Omega^2\bigg{]}\,, \\
\Phi(r,t)&=&\pm\frac{1}{4 \pi }\ln \left[ \left(1-\frac{2 c}{r}\right)^{\frac{\alpha}{\sqrt{3}}}(at+1)^{\sqrt{3}}\right]\,,
\end{eqnarray}
where  $\alpha=\pm\frac{\sqrt{3}}{2}$.
From the Ricci scalar
\begin{equation}\label{cur}
{\mathcal R}=\frac{12 ca^2(r-c)-3a^2 r^2}{2 r^2(at+1)^3}\left(1-\frac{2c}{r}\right)^{-2-\alpha}+\frac{2c^2(1-\alpha^2)}{(a t+1)r^4}
\left(1-\frac{2c}{r}\right)^{-2+\alpha}\,,
\end{equation}
we see that the curvature singularities are located at $r =2c$ (timelike singularity) and $ t=-1/a$(spacelike singularity).
Using $\Theta_\pm$ to label the expansion of  null geodesics, we have
\begin{eqnarray}
\Theta_+&=&\frac{1}{\sqrt{h}}\dfrac{\partial \sqrt{h}}{\partial \lambda^+}\,,\label{equ:188}\\
\Theta_-&=&\frac{1}{\sqrt{h}}\dfrac{\partial \sqrt{h}}{\partial \lambda^-}\,,\label{equ:189}
\end{eqnarray}
where $\sqrt{h}=R^2\sin\theta=(at+1)r^2\left(1-\frac{2c}{r}\right)^{1-\alpha}\sin\theta$, and $\lambda_{\pm}$ is the affine parameter of the null geodesics. The tangent to the null geodesic with affine parameter is
\begin{eqnarray}
\dfrac{\partial}{\partial \lambda^+}&=&\dfrac{1}{\left(at+1\right)}\left(1-\frac{2c}{r}\right)^{-\alpha}\dfrac{\partial}{\partial t}+\dfrac{1}{\left(at+1\right)}\dfrac{\partial}{\partial r}\,,\\
\dfrac{\partial}{\partial \lambda^-}&=&\dfrac{1}{\left(at+1\right)}\left(1-\frac{2c}{r}\right)^{-\alpha}\dfrac{\partial}{\partial t}-\dfrac{1}{\left(at+1\right)}\dfrac{\partial}{\partial r}\,.
\end{eqnarray}
We can get
\begin{gather}\label{expansion}
\Theta_+=\left\lbrace ar^2\left(1-\frac{2c}{r}\right)^{-\alpha}+\left[2r-\frac{2cr(\alpha-1)}{r-2c}\right](at+1)\right\rbrace r^{-2}(at+1)^{-2}\,,\\
\Theta_-=\left\lbrace ar^2\left(1-\frac{2c}{r}\right)^{-\alpha}-\left[2r-\frac{2cr(\alpha-1)}{r-2c}\right](at+1)\right\rbrace r^{-2}(at+1)^{-2}\,.
\end{gather}
The apparent horizon  satisfies  $\Theta_+=0,\Theta_-<0$, which is located at
\begin{equation}\label{ra}
\frac{a}{at_{AH}+1}=-\frac{2}{r_{AH}^2}\left[ r_{AH}-c(1+\alpha)\right](1-\frac{2 c}{r_{AH}})^{\alpha-1}\,.
\end{equation}
The detailed analyses about the HMN sapcetime can be found in \cite{bibitem56}. We shall focus on the case $-\infty\leq t \leq -\frac1a $ and $a<0$ because it corresponds to a black hole solution. When $a\neq 0$ and $\alpha =\frac{\sqrt{3}}{2}$, the apparent horizon in the HMN spacetime has "S-curve" shape. However, this does not make differences in our results. Therefore, in this paper, we just study the case $\alpha=-\frac{\sqrt{3}}{2}$ for simplicity. The results remain true when $\alpha=\frac{\sqrt{3}}{2}$.
The Misner\-Sharp mass\cite{msm} is defined by
 \begin{equation}\label{equ:5}
 M=\frac{R}{2}\left(1-g^{ab}\nabla_a R\nabla_bR\right)\,.
 \end{equation}
where $R$ denotes the areal radius. From \cite{bibitem57}, we know that in spherically symmetric spacetimes, the apparent horizon satisfies
\begin{equation}
g^{ab}\nabla_a R\nabla_bR=0\,.
\end{equation}
Therefore, on the apparent horizon, the Misner-sharp mass  becomes
\begin{eqnarray}\label{equ:58}
M&=&\frac{R}{2}\,.
\end{eqnarray}
\section{Critical behaviour of HMN scalar field with conformal flatness ($c\neq 0$) }\label{cfcc}
\subsection{Matching at a timelike  boundary}
\begin{figure}[htbp]
\centering
\includegraphics[width=6.96cm,height=4.70cm]{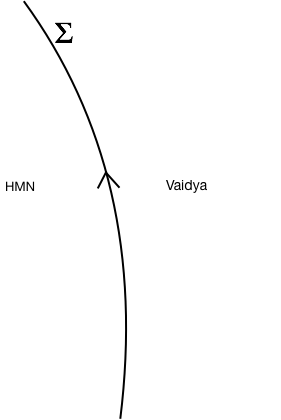}
\caption{$a\neq0, c=0$ timelike hypersurface}
\label{Fig5}
\end{figure}

To study the critical phenomenon of HMN massless scalar field, we start with the simple case $c=0$, where the spacetime is conformally flat\cite{bibitem1}. First, we need to join the HMN solution with an outgoing Vaidya solution such that the resulting spacetime is asymptotically flat. In this section, we assume that the boundary connecting the two solutions is a timelike hypersurface.  We shall use "$-$" to label the inner HMN spacetime and  "$+$"  to label the exterior Vaidya spacetime (see \fig{Fig5}).

The interior spacetime is described by \eq{m2} for $c=0$
\begin{eqnarray}\label{equ:131}
ds_{-}^2=(1+at)\left[-dt^2+dr^2+r^2 \left(d\theta^2+\sin\theta^2 d\phi^2\right)\right]\,.
\end{eqnarray}
The exterior spacetime is described by the outgoing Vaidya metric

\begin{equation}\label{equ:132}
ds_{+}^2= -f_+ dU^2-2 dUdR+R^2 d\Omega^2\,,
\end{equation}
where $f_+=1-\frac{2 m(U)}{R}$.  We choose $\xi^i=\{\lambda,\theta,\phi\}$ as the  intrinsic coordinates on the hypersurface. $\Sigma$ is determined by functions $t(\lambda), r(\lambda)$ from the interior and $T(\lambda), R(\lambda)$ from the exterior. The induced metric on $\Sigma$ is given by
\begin{eqnarray}\label{equ:211}
ds_{-}^2\Big{|}_{\Sigma}&=&(a t+1)\left[(-\dot{t}^2+\dot{r}^2)d\lambda^2+r^2 \left(d\theta^2+\sin\theta^2 d\phi^2\right)\right]\,,\\\label{equ:208}
ds_{+}^2\Big{|}_{\Sigma}&=&-\left(f_+\dot{U}+2 \dot{R}\right) \dot{U}d\lambda^2+R^2 d\Omega^2\,.
\end{eqnarray}
Here the do``$.$'' means the derivative with respect to $\lambda$.

We use the Darmois junction conditions to match the solutions across $\Sigma$
\begin{eqnarray}\label{equ:126}
ds_{-}^2\Big{|}_{\Sigma}&=&ds_{+}^2\Big{|}_{\Sigma}\,,\\ \label{equ:209}
k^{-}_{a b}\Big{|}_{\Sigma}&=&k^{+}_{a b}\Big{|}_{\Sigma}\,,
\end{eqnarray}
where $k_{ab}$ is the extrinsic curvature of $\Sigma$.

Denote the coordinates of the four-dimensional spacetime by $\{x^\mu\}$. $\Sigma$ is determined by the functions $\{x^\mu(\xi^i)\}$. Then  the components of $k_{ab}$ can be calculated from
\bean
k_{ij}=-n_\mu\frac{\partial^2 x^\mu}{\partial\xi^i\xi^j}-n_\nu\Gamma^\nu_{\mu\rho}\pp{x^\mu}{\xi^i}\pp{x^\rho}{\xi^j}\label{ktime}
\eean
where $n_a$ is the spacelike normal to $\Sigma$. Computing $k_{ij}$ from the interior and exterior, respectively, we obtain the nonvanishing components

\begin{eqnarray}\label{equ:127}
k^{\mathcal{-}}_{\lambda\lambda}&=&\dfrac{a\dot{r}(\dot{r}^2-\dot{t}^2)-2 (at +1)(\dot{t}\ddot{r}-\dot{r}\ddot{t})}{2\sqrt{(a t +1)(\dot{t}^2-\dot{r}^2)}}\,,\\ \label{equ:133}
k^{\mathcal{+}}_{\lambda\lambda}&=&\dfrac{\dot{U}}{\left(f_+ \dot{U}^2+2 \dot{R} \dot{U}\right)^{\frac12}}\left[-\dfrac{\dot{U}^2}{R}\left(\dfrac{m f_+}{R}-\dfrac{d m}{d U}\right)+\dot{R}\left(\dfrac{\ddot{U}}{\dot U}-3\dfrac{\dot{U} m}{R^2}\right)-\ddot{R}\right] \,,\\ \label{equ:134}
k^{\mathcal{-}}_{\theta\theta}&=&\dfrac{1}{\sin^2\theta k^{\mathcal{I}}_{\phi\phi}}=\dfrac{a r^2 \dot{r}+2 (at +1)\dot{t}r}{2\sqrt{(a t+1) (\dot{t}^2-\dot{r}^2)}}\,,\\\label{equ:135}
k^{\mathcal{+}}_{\theta\theta}&=&\dfrac{1}{\sin^2\theta k^{\mathcal{E}}_{\phi\phi}}=\dfrac{R}{\left(f_+\dot{U}^2+2\dot{R}\dot{U}\right)^{\frac{1}{2}}}
\left(\dot R+f_+\dot U\right)\,.
\end{eqnarray}
Substituting  \eqs{equ:211} and \meq{equ:208} into \eq{equ:126}, we have
\begin{eqnarray}\label{equ:215}
R&=&\sqrt{a t+1} r\,,\\\label{equ:216}
(at+1)(\dot{t}^2-\dot{r}^2)&=&\left(f_+\dot{U}+2 \dot{R}\right) \dot{U}\,.
\end{eqnarray}
Substituting \eqs{equ:134} and \meq{equ:135} into \eq{equ:209}, with the help of \eq{equ:216}, we obtain
\begin{equation}\label{equ:217}
f_+ \dot{U}+ \dot{R}=\dfrac{a r\dot{r}+2(a t+1) \dot{t}}{2\sqrt{a t+1}}\,.
\end{equation}
\eq{equ:215} yields
\begin{equation}\label{equ:218}
\dot{R}=\dfrac{a r\dot{t}}{2\sqrt{(at+1)}}+\sqrt{a t+1}\dot{r}\,.
\end{equation}
Therefore, one can solve \eqs{equ:216}-\meq{equ:218} and obtains
\bean
\dot{U}\eqn \dfrac{2(a t+1)^\frac32\left(\dot{t}-\dot{r}\right)}{ar+2 (at +1)}\,. \\ \label{equ:219}
f_+\eqn1-\frac{m}{R}= 1-\dfrac{a^2r^2}{4 (at +1)^2}\,.
\eean
Thus,
\begin{equation}
m=\dfrac{a^2r^3}{8(at+1)^{\frac32}}\,.
\end{equation}
To proceed, we  calculate the following derivatives:
\begin{eqnarray}\label{equ:221}
\ddot{U}&=&\frac{\sqrt{a\hat{t}}}{(r+2\hat{t})^2}\left[2\hat{t} \dot r^2-(4\hat{t} +3r)\dot t\dot r+2\hat{t} (2\hat{t}  \ddot t-2\hat{t} \ddot r+\dot t^2)+r(2\hat{t} \ddot t-2\hat{t} \ddot r+3\dot t^2) \right]  \,, \\
\ddot{R}&=& \sqrt{a\hat{t} }\ddot r+\frac{2\sqrt{a}\hat{t} \dot t\dot r+\sqrt{a}(2\hat{t} \dot t\dot r+2\hat{t} r\ddot t-r\dot t^2)}{4\hat{t} ^{3/2}}\,, \\
\dfrac{d m(U)}{d U}&=&\frac{\dot m}{\dot U}=\frac{3r^2(r+2\hat{t} )(r\dot t-2\hat{t} \dot r)}{32\hat{t} ^4(\dot r-\dot t)}  \,.
\end{eqnarray}
where $\hat{t}=t+a^{-1}$.
Substitute the above results into \eq{equ:133} and according to \eq{equ:209},let the right-hand side of \eq{equ:127}  be equal to the right-hand side of \eq{equ:133}. After a lengthy calculation, we obtain the following result, which is surprisingly simple
\bean
(\dot t-\dot r)(\dot t+\dot r)^2=0
\eean
Obviously, the solution is $\dot{r}=\dot{t}$ or $\dot{r}=-\dot{t}$. But this means that the hypersurface is null, inconsistent with our assumption. Therefore, we conclude that the two spacetimes cannot be matched through a timelike hypersurface if the continuity of the extrinsic curvature is required.

\subsection{Matching at  a null hypersurface }\label{conformal}
Matching the two solutions at a null hypersurface is more complicated than at a timelike hypersurface.  We shall follow the method in \cite{bibitem5} and \cite{bibitem11}. First we use the coordinate transformation  $v=t+r$ to replace the coordinate $r$ in \eq{m2} and obtain the metric in the interior
\begin{equation}\label{m2a}
ds^2=-\left[a\left(v-r\right)+1 \right]dv(dv-2dr)+R^2 \left(d\theta^2+\sin^2\theta d\phi^2\right)\,.
\end{equation}
where
\begin{gather}
R^2=\left[a (v-r)+1\right] r^2\,.\label{equ:128}
\end{gather}
Let $\Sigma$ be the null hypersurface $v=v_0$.
The normal to $\Sigma$ is
\begin{equation}
n_a^-=s^{-1} dv_a\,,
\end{equation}
where $s$ is a negative arbitrary function such that $n^a_-$ is a future directed vector. We can introduce a transverse null vector $N_a$ by requiring
\begin{eqnarray}\label{equ:66}
n_a N^a&=&-1\,,\\
N^a N_a&=&0\,.
\end{eqnarray}
Without loss of generality, we assume that $N_a^-=N_v dv_a+N_r dr_a$. Then it is easy to show that
\begin{equation}\label{equ:140}
N_a^-=s [a\left(v-r\right)+1]\left(\frac{1}{2}dv_a- dr_a\right)\,.
\end{equation}
Now we choose $s=-\dfrac{1}{a (v-r)+1}\dfrac{\partial R}{\partial r}$.
Choose $\xi^i=\{R,\theta,\phi\}$ to be the  intrinsic coordinates on $\Sigma$. Then $\Sigma$ can be determined by
\bean
v= v_0 \,,\ \
r=r(R)\,,\ \
\theta=\theta\,, \ \
\phi= \phi \,,
\eean
where $r(R)$ is determined by \eq{equ:128} with $v=v_0$. We define $e_{(a)}^{-\mu}\equiv\dfrac{\partial x^{\mu}_{-}}{\partial \xi^a}$ as given in the\cite{bibitem5}. Thus
\begin{eqnarray}\label{e11}
e_{(1)}^{-r}=\frac{2 r R}{2 R^2-a r^3}\,,\label{e12}\ \
e_{(2)}^{-\theta}=1\,,\label{e13}\ \
e_{(3)}^{-\phi}=1\,.
\end{eqnarray}
Similarly to \eq{ktime}, the transverse extrinsic curvature for the null surface is given by \cite{bibitem11}.
\begin{eqnarray}\label{equ:130}
k_{ij}=-N_\mu\frac{\partial^2 x^\mu}{\partial\xi^i\xi^j}-N_\nu\Gamma^\nu_{\mu\rho}\pp{x^\mu}{\xi^i}\pp{x^\rho}{\xi^j}
\end{eqnarray}
By straightforward calculation, we find
\begin{equation}\label{equ:199}
k_{ab}^-= \dfrac{-3 a^2r}{\left(3ar-2av_0-2\right)^2\sqrt{a\left(v_0-r\right)+1}}dR_a dR_b+\dfrac{r\left[a\left(r-2v_0\right)-2\right]\left(3ar-2av_0-2\right)}{8\left[a\left(v_0-r\right)+1\right]^{\frac32}}\left (d\theta_a d\theta_b+\sin\theta^2d\phi_a d\phi_b\right)\,.
\end{equation}

\begin{figure}[htbp]
\centering
\includegraphics[width=6cm,height=3.6cm]{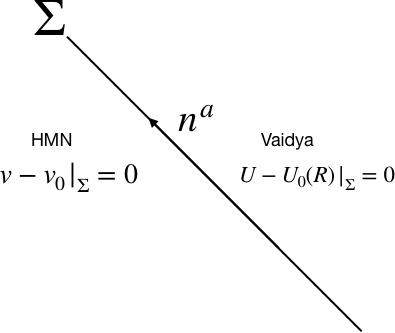}
\caption{ingoing null hypersurface $v-v_0=0$ or $U-U_0(R)=0$}
\label{Fig6}
\end{figure}

On the other hand, we need to join the HMN solution to the exterior Vaidya spacetime \meq{equ:132} at $\Sigma$, as shown in \fig{Fig6}. Assume that the null surface $\Sigma$ can be described by   $U=U_0(R)$ from the Vaidya solution. It follows from \meq{equ:132} that
\begin{equation}\label{equ:299}
\dfrac{dU_0}{dR}=-\frac{2R}{R-2m[U_0(R)]} \,.
\end{equation}
The spacetime coordinates $\{x_+^\mu\}$ can be expressed as functions of $\xi^i$:
\bean
U= U_0(R)\,,\ \
R= R\,, \ \
\theta= \theta \,, \ \
\phi= \phi \,.
\eean
Define $e_{(i)}^{+\mu}\equiv\dfrac{\partial x^{\mu}_{+}}{\partial \xi^i}$   and it is easy to find
\begin{eqnarray}\label{b1}
e_{(1)}^{+U}=-\dfrac{2}{f_{+}}\,,\ \ \label{b2}
e_{(1)}^{+R}=e_{(2)}^{+\theta}=e_{(3)}^{+\phi}=1\,.
\end{eqnarray}
The normal to $\Sigma$ is given by
\begin{equation}
n_{a}^+=\beta^{-1}\nabla_a ( U-U_0(R))=\beta^{-1}\left(dU_a+\dfrac{2}{f_+}dR_a\right)\,.
\end{equation}
where $\beta$ is a negative function which will be determined later.
Then the transverse null vector $N^a$ in \eq{equ:66} is
\begin{equation}\label{en}
N_a^+=\dfrac{\beta f_+}{2}dU_a\,,
\end{equation}
 The continuity condition on $\Sigma$ requires  \cite{bibitem11}
\begin{equation}\label{equ:120}
N_\mu^+ e^{+\mu}_{(i)}\Big{|}_{v=v_0}=N_\mu^- e^{-\mu}_{(i)}\Big{|}_{v=v_0}\,.
\end{equation}
This also guarantees that the normal vectors $n^a$ defined on both sides are the the same. According to \eqs{equ:140},\meq{e11}-\meq{e13}, \eqs{b1}, \meq{en}, we find that the nontrivial equations  in \eq{equ:120} are
\begin{eqnarray}
N_\mu^- e^{-\mu}_{(1)}\Big{|}_{v=v_0}&=&N_r^-e_{(1)}^{-r}=1\,,\\
N_\mu^+ e^{+\mu}_{(1)}\Big{|}_{v=v_0}&=&N_U^+e_{(1)}^{+U}=-\beta\,.
\end{eqnarray}
Hence,  we get $\beta=-1$.

Now we can calculate the corresponding transverse extrinsic curvature from \eqs{equ:130} and obtain
\begin{equation}\label{equ:145}
k_{ab}^+=\dfrac{-2m'(r)r'(R)}{\sqrt{a\left(v_0-r\right)+1} r-2m(r)}dR_adR_b+\left(\frac{R}{2}-m[U_0(R)]\right)\left(d\theta_a d\theta_b+\sin^2\theta d\phi_a d\phi_b\right)\,.
\end{equation}
Since $k_{ab}^+=k_{ab}^-$,    \eqs{equ:199} and  \meq{equ:145} give rise to
\begin{equation}
k_{RR}^+\big{|}_{\Sigma}=k_{RR}^-\big{|}_{\Sigma}\,.
\end{equation}
By integration, we obtain $m(r)$ as
\begin{eqnarray}
m(r)=\frac{-8 + a^3\left(-27 r^3 + 12 r v_0^2 - 8 v_0^3\right) + 12 a \left[r-2 (v_0+ 18c_1 )\right] + 24 a^2 \left[-v (v_0 + 18 c_1) + r (v_0+ 27 c_1)\right]}{-216 a \left[1 + a (v_0-r)\right]^\frac{3}{2}}\,,
\end{eqnarray}
where $c_1$ is an integral constant. Using
$k_{\theta\theta}^+\big{|}_{\Sigma}=k_{\theta\theta}^-\big{|}_{\Sigma}$,
we can fix $c_1$:
\begin{equation}
c_1=\frac{-(av_0+1)^2}{54 a}\,.
\end{equation}
Consequently,
\begin{eqnarray}\label{equ:125}
m(r)=\dfrac{a^2 r^3}{8 \left[a\left(v_0-r \right)+1\right]^{3/2}}\,.
\end{eqnarray}
Note that
\begin{equation}
m(r)\big{|}_{\Sigma}=m(U(R(r))\big{|}_{\Sigma}\,.
\end{equation}
So, \eq{equ:125} together with \eq{equ:299} specifies a unique metric function $m(U)$ in the Vaidya solution. Therefore, we have matched the conformally flat spacetime with the Vaidya spacetime at the null hypersurface.

\subsection{Mass of the black hole }  \label{masscf}
From \eq{equ:5},
one can calculate the Misner-Sharp mass for the HMN spacetime described by metric \meq{equ:131} and obtain
\begin{equation}\label{equ:6}
M=\frac{r^3 a^2}{ 8 \left(a t+1\right)^{\frac{3}{2}}}\,.
\end{equation}
Therefore,  $m(r)$ in \eq{equ:125} is just the Misner-Sharp mass at $v=v_0$.
The apparent horizon determined by  \eq{ra} takes the simple form for $c=0$:
\begin{eqnarray}
\frac{a}{a t_{AH}+1}=-\frac{2}{r_{AH}}\,.  \label{tr}
\end{eqnarray}
Note that the null surface is determined by
\begin{equation}
v_0=t+r\,,  \label{ns}
\end{equation}
\eqs{tr} and \meq{ns} immediately gives the coordinates at  the intersection of $\Sigma$ and the apparent horizon:

\begin{eqnarray} \label{tr2}
r_i&=&2v_0+\frac2a,\\\label{ns2}
t_i&=&-v_0-\frac2a\,.
\end{eqnarray}
Since $r>0$ and $a<0$, from \eq{tr2}, we see that the existence of the intersection requires
\bean
v_0>|a|^{-1}
\eean

\begin{figure}[htbp]
\centering
\includegraphics[width=6.6cm,height=7.2cm]{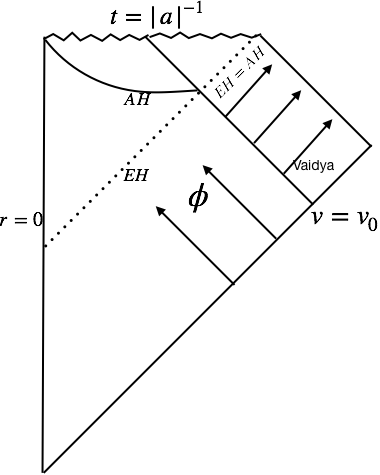}
\caption{Penrose diagram for HMN spacetime ($a\neq0, c=0$) matching with an outging Vaidya spacetime. The singularity is located at $t=-a^{-1}$. The HMN spacetime matches with the outgoing Vaidya spacetime at $v=v_0$. }
\label{Fig1}
\end{figure}

Therefore, the Misner-Sharp  mass at the intersection  is
\begin{equation}\label{mass4}
M_i=\sqrt{-a}\left(v_0+\frac{1}{a}\right)^{\frac32}\,.
\end{equation}
 As is known, the event horizon coincides with the apparent horizon in the outgoing Vaidya spacetime as shown in \fig{Fig1}. It is also known that the mass function $m$ in the Vaidya metric is constant alone the event horizon \cite{vaidyaevent}. Thus, it is natural to take the mass in \eq{mass4}  to be the mass of the black hole.

To investigate the critical behavior as $a\rightarrow 0$, we impose the condition that $r_i$ in \eq{ns2} does not change with $a$. This means that $v_0$ must take the form
\bean
v_0=V-\frac{1}{a}
\eean
where $V$ is a positive constant independent of $a$. Thus, \eq{mass4} gives the mass of black hole:
\begin{equation}\label{mass3}
M_{bh}=\sqrt{-a}V^{\frac32}\,.
\end{equation}
\eq{mass3} shows that the mass of black hole can be put in the form of \eq{sform} and the scaling exponent is $\gamma = 0.5$. Obviously, as $a$ approaches zero, the mass of the black hole vanishes and
the spacetime becomes Minkowski.

\section{ Collapse of the general HMN scalar field}\label{gscc}
In this section, we shall investigate the gravitational collapse associated with a general HMN scalar field ($ a \neq 0, c \neq 0$).
\subsection{Matching to an outgoing Vaidya solution at a null hypersurface }
\begin{figure}[htbp]
\centering
\includegraphics[width=6.6cm,height=7.2 cm]{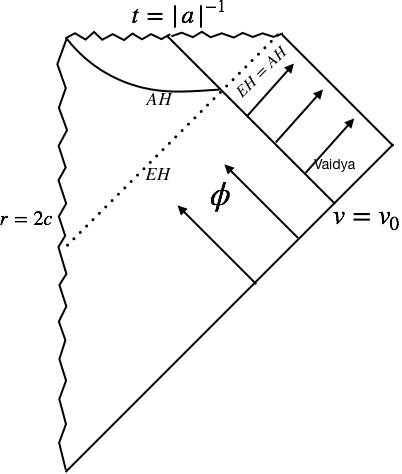}
\caption{Penrose diagram for the HMN solution($a\neq0, c\neq0,\alpha=-\frac{\sqrt{3}}{2}$) matching with an outging Vaidya solution. There are two singularities at $t=|\frac1a$| and $r=2c$, where $r=2c$ is a naked singularity. }
\label{2}
\end{figure}

Under the coordinate transformation
\bean
v=t+h(r)
\eean
 \eq{m2} can be rewritten in the form
 \begin{equation}\label{inme}
 ds^2=-\left[a(v-h(r))+1\right]\left(1-\frac{2c}{r}\right)^{\alpha}\left( dv^2-2h'(r) dv dr\right)+a\left[(v-h(r))+1\right]\left(1-\frac{2c}{r}\right)^{1-\alpha}r^2 d\Omega^2\,.
 \end{equation}
Here, the function $h(r)$ satisfies
\begin{eqnarray}
h'(r)=\left(1-\frac{2 c}{r}\right)^{-\alpha}\,.
\end{eqnarray}
The areal radius $R$ takes the form
 \begin{equation}\label{equ:143}
 R=\sqrt{a(v-h(r))+1}\left(1-\frac{2c}{r}\right)^{\frac{1-\alpha}{2}}r\,.
 \end{equation}
Similarly to section 3, we match the solution with an outgoing Vaidya solution at the null hypersurface
$v=v_0$ (see \fig{2}). Substitution of $v=v_0$ into \eq{equ:143} yields the function $r=r(R)$. By the method in section \ref{conformal}, the extrinsic curvature can be calculated as

\begin{equation}\label{equ:146}
\begin{split}
k_{ab}^- &=\left( \dfrac{a h'(r)r'(R)}{1+a(v_0-h(r))}-\dfrac{r''(R)}{r'(R)}\right)dR_a dR_b -\bigg{[}\dfrac{(2+\sqrt{3})c-2r+4ac (1 - 2 c/r)^{-\frac{\sqrt{3}}{2}} r - 2a (1 - 2 c/r)^{-\frac{\sqrt{3}}{2}} r^2 }{4\left[1+a(v_0-h(r))\right]r'(R)}\\
&+\frac{((2 +\sqrt{3}) c-2 r)a h(r)+2ar v_0 +a r (2 c - r) h'(r)-(2+\sqrt{3})ac v_0 }{4\left[1+a(v_0-h(r))\right]r'(R)}\bigg{]}\left( d\theta_a d\theta_b+\sin^2\theta d\phi_a d\phi_b\right)\,.
\end{split}
\end{equation}
where
\begin{eqnarray}\label{equ:145aa}
r'(R)&=&-\dfrac{2(1-\frac{2c}{r})^{\frac{\alpha-1}{2}}(r-2 c)\sqrt{a\left[v_0-h(r)\right]}}{\left((2 + \sqrt{3}) c - 2 r\right)(1+av_0) - a((2 +\sqrt{3})c-2 r) h(r) + a r(-2 c + r)h'(r)}\,,\\\notag
r''(R)&=&-\Big{[}((2+\sqrt{3}) c - 2 r)(1+av_0)-a((2 + \sqrt{3}) c - 2 r) h(r)+ ar (-2 c + r)h'(r)^3\Big{]}^{-3}\\\notag
&+&\Big{\lbrace}2(1-2 c/r)^{\frac{\sqrt{3}}{2}}\big{[}a^2c^2 h(r)^2 +  2 a((2 + \sqrt{3}) c - 2 r) (2 c - r) r (1+av_0) h'(r) + a^2r^2 (-2 c + r)^2 h'(r)^2 \\\notag
&-&2ah(r)(c^2(1+a v_0) + a((2 + \sqrt{3}) c - 2 r) (2 c - r) r h'(r)+ ar^2 (-2 c + r)^2 h''(r)) \\
&+&(1+a v_0) (c^2(1+a v_0) + 2 ar^2 (-2 c + r)^2 h''(r))\big{]}\Big{\rbrace}\,.
\end{eqnarray}
It is easy to see that $k_{ab}^+$ takes the same form as \eq{equ:145}. Then $k_{ab}^-=k_{ab}^+$ yields

\begin{equation}\label{equ:100}
\begin{split}
m(r)&=\dfrac 12 (1 - 2 c/r)^{(2 -\sqrt{3})/4}r \sqrt{a(v_0-h(r))+1}
+\frac{1}{4(a(v_0-h(r))+1)r'(R)}\Big{(}(2+\sqrt{3})c-2r\\
&+4acr(1 - 2 c/r)^{-\frac{\sqrt{3}}{2}}-2a(1 - 2 c/r)^{-\frac{\sqrt{3}}{2}}r^2+((2 +\sqrt{3}) c - 2 r) a v_0 \\\notag
&-a((2 + \sqrt{3}) c - 2 r) h(r) +a r (-2 c + r) h'(r)\Big{)}\,.
\end{split}
\end{equation}
 Thus, by our construction, the metric of the resulting spacetime is continuous and the extrinsic curvature of the null hypersurface  is also continuous. Therefore, we have shown that the general HMN spacetime ( $a\neq0,c\neq0$) and Vaidya spacetime can be matched at a null hypersurface as showed in  \fig{2}.

\subsection{ Mass of the black hole}
By the argument in section \ref{masscf}, one can show that $m(r)$ in \eq{equ:100} is exactly the Misner-sharp mass for the metric in \eq{inme}.  Therefore, from \eqs{ra},\meq{equ:58} and \meq{equ:143}, we can obtain the mass on the apparent horizon $m_{AH}(r)$:
\begin{equation}\label{mass11}
m_{AH}(r)=\frac{R_{AH}(r)}{2}=\sqrt{-a}r^2[8r-8c(1+\alpha)]^{-\frac12}\left(1-\frac{2c}{r}\right)^{-\alpha+1}\,.
\end{equation}
The ingoing null hypersurface boundary is defined by
\begin{equation}\label{equ:16}
v_0=t+h(r)\,.
\end{equation}
From the \eq{ra} and the \eq{equ:16},  we can get the coordinates $(r_i,t_i)$ at the intersection of the apparent horizon and the null hypersurface $v=v_0$, which satisfies

\begin{figure}[htbp]
\centering
\includegraphics[width=6.6cm,height=7.2cm]{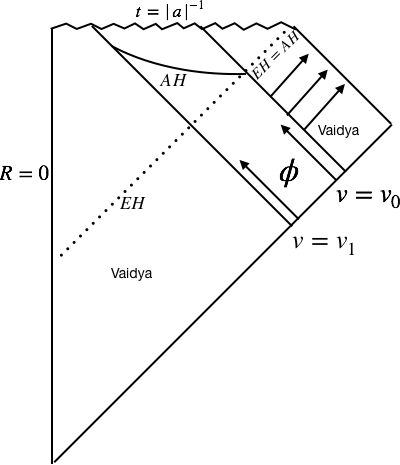}
\caption{Penrose diagram for the HMN scalar field ($a\neq0, c\neq0$) matching with two outgoing Vaidya spacetimes.
We see that the naked singularity in \fig{2} has been replaced by the Vaidya spacetime.
 }
\label{3}
\end{figure}

\bean\label{inse11a}
v_0+\frac1a\eqn\frac{(1 - 2 c/r_i)^{\frac{\sqrt{3}}{2}}(2 c-r_i) r_i + \left[(-2+\sqrt{3})c+2r_i\right]h(r_i)}
{(-2+\sqrt{3})c+2r_i}\\
at_i+1\eqn\frac{a(1-2 c/r_i)^{\frac{\sqrt{3}}{2}}r_i(2c-r_i)}{-2c +\sqrt{3}c+2r_i}>0\,.
\eean
Similarly, we choose the null hypersurface which intersects with the apparent horizon at a fixed radius  $r_i=r_0$, i.e., independent of $a$. Again, we take the Misner-Sharp mass at the intersection as the black hole mass. Then,  \eq{mass11} gives the mass of the black hole
\bean
M_{bh}(r_0)=\sqrt{-a}f(r_0,c)\,, \label{mbhr}
\eean
where
\bean
f(r_0,c)=r_0^2[8r_0-8c(1+\alpha)]^{-\frac12}\left(1-\frac{2c}{r_0}\right)^{-\alpha+1}\,.
\eean
\eq{mbhr} shows clearly that the black hole mass satisfies the power law with $\gamma=0.5$. However, the spacetime for $a<0$ possesses a naked singularity $r=2c$ (see \fig{2}), in violation of the cosmic censorship conjecture. To remove the naked singularity, we join another outgoing Vaidya spacetime at $v=v_1(v_1<v_0)$, as shown in \fig{3}. No naked singularity exists in this new spacetime.

When we study the relation between the mass and the parameter $a$, we treat $c$ as a constant. We see that $M_{bh}\rightarrow 0$ as $a\rightarrow 0$. When $a=0$, there is no black hole but a naked singularity as shown in section \ref{sec-cs}.

\subsection{  "Critical spacetime": $a=0$ and $c\neq 0$  } \label{sec-cs}
Now we study the critical HMN spacetime. For  $a=0$,  \eq{m2} becomes
\begin{eqnarray}
ds^2=-\left(1-\frac{2c}{r}\right)^{\alpha} dt^2+\left(1-\frac{2c}{r}\right)^{-\alpha}dr^2+r^2\left(1-\frac{2c}{r}\right)^{1-\alpha}d\Omega^2\,,
\end{eqnarray}
To calculate the apparent horizon of the spacetime, we first choose two families of radial null vector fields
\bean
\dfrac{\partial}{\partial \lambda^+}\eqn\left(1-\frac{2c}{r}\right)^{-\alpha}\dfrac{\partial}{\partial t}+\dfrac{\partial}{\partial r}\\
\dfrac{\partial}{\partial \lambda^-}\eqn\left(1-\frac{2c}{r}\right)^{-\alpha}\dfrac{\partial}{\partial t}-\dfrac{\partial}{\partial r}\,.
\eean
where $\lambda^\pm$ is the affine parameter of the null geodesic.

\begin{figure}[htbp]
\centering
\includegraphics[width=5cm,height=6.6 cm]{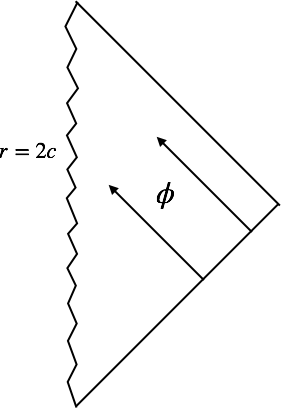}
\caption{Penrose diagram for the HMN spacetime($a=0, c\neq0$). There is a naked singularity at $r=2c$.}
\label{Fig4}
\end{figure}

According to \eq{expansion}
\begin{eqnarray}
\Theta_+&=&-\dfrac{2\left(c-r+c\alpha\right)}{r(-2 c+r)}\,,\\
\Theta_-&=&\dfrac{2(c-r+c\alpha)}{r(-2 c+r)}\,.
\end{eqnarray}
Therefore, the spacetime has no apparent horizon when  $r>2c$. According to \eq{cur}, the spacetime possesses a naked singularity  at $r=2 c$ (see \fig{Fig4}).

Since the spacetime is asymptotically flat,  we calculate the ADM mass and find
\begin{equation}\label{mass3b}
M=c \alpha\,.
\end{equation}
When $a\neq0$, the mass of the black hole  is given by \eq{mbhr}, which shows clearly that $M\rightarrow 0$ as $a\rightarrow 0$. However, when $a=0$, as we just discussed, the spacetime is not Minkowski and its ADM mass is nonzero. Therefore, there exists a mass gap between the black hole solution ($a\neq 0$)  and its limiting spacetime ($a=0$).

\section{ Conclusion} \label{sec-con}
In this paper, we have used the "cut and paste" method to construct  analytical models  and  study the critical phenomena of the HMN scalar filed. We have shown that the HMN solution with conformal flatness ($c\neq 0$) can be matched with the Vaidya solution along a null hypersurface, but not a timelike hypersurface. We have derived the differential equation which specifies the metric function in the Vaidya solution. For  $c\neq0$, we have joined the scalar field onto two pieces of Vaidya spacetimes to avoid the naked singularity.

We have studied the gravitational collapse for the HMN scalar field and shown that black hole mass satisfies the power law with  $\gamma=0.5$. This is consistent with  previous results in the literature. When $c\neq0$, the HMN spacetime has no CSS  and the black hole also turns on at infinitely small mass. The result is different from the model in \cite{bibitem2}, which shows that the formation of black holes may turn on at finite mass when the gravitational collapse has no self-similarity. On the other hand, the mass gap exists between the black hole and the naked singularity during the gravitational collapse of HMN scalar field when $c\neq 0$ as discussed in  section \ref{gscc}. Our work suggests that critical collapse can be studied from analytical models which are constructed by known solutions. More models should be investigated in the future in order to test the universal features in gravitational collapse.

\appendix
\section{Self-similarity of HMN spacetime}\label{appendix}

In this appendix, we will prove that the HMN spacetime is CSS (continuous self similar) only when $c=0$ and $a\neq 0$. A spacetime is continuous self-similar if there exists a conformal Killing vector field $\xi^{a}$ satisfying
\begin{equation}\label{sel}
\nabla_{(a}\xi_{b)}=g_{ab}\,.
\end{equation}
As a result of the spherical symmetry, we can write
\begin{equation}
\xi^a=x\ppa{t}{a}+y\ppa{r}{a}\,.
\end{equation}
Here, $x$ and $y$ are  functions of $r$ and $t$.
Substituting this expression into \eq{sel}, we find that
\begin{gather}\label{sim}
y R_{,r}+x R_{,t}=R\,,\\ \label{sim1}
y \nu_{,r}+ x \nu_{,t}+y_{,r}=1\,, \\ \label{sim2}
y \lambda_{,r}+x \lambda_{,t}+x_{,t}=1 \,,\\ \label{sim3}
m  y_{,t}-n x_{,r}=0\,.
\end{gather}
Here,
\begin{gather}\label{equ:165}
R^2= \left(a t + 1\right) r^2  \left(1 -\frac{2c}{r}\right)^{1 - \alpha}\,, \notag\\
\lambda =  \frac{1}{2}\log\left[(a t +1) \left(1 -\frac{2c}{r}\right)^\alpha\right]\,, \notag\\
\nu = \frac{1}{2}\log\left[(a t + 1)  \left(1 -\frac{2c}{r}\right)^{-\alpha}\right] \,,\notag\\
m = (a t + 1)  \left(1 -\frac{2c}{r}\right)^{-\alpha}\,, \notag\\
n = (a t + 1)  \left(1 -\frac{2c}{r}\right)^{\alpha}\,.
\end{gather}
From \eq{sim}, we can get
\begin{equation}\label{sim4}
x=-\frac{1}{ar(-2 c+r)}(1+a t)(4 c r -2 r^2-2c y+\sqrt{3} cy+2 ry)\,.
\end{equation}
Substituting \eq{sim4} into \eq{sim1}, we obtain
\begin{equation}\label{sim5}
y=\sqrt{r}\sqrt{-2c+r}D(t)\,,
\end{equation}
where $D(t)$ is a integration function of $t$.
Putting \eqs{sim4} and \meq{sim5} into \eq{sim3}, we find
\begin{equation}\label{sim6}
-\frac{c(1-\frac{2c}{r})^{-\sqrt{3}}((-2+\sqrt{3})c-\sqrt{3}r)}{a(-2c+r)^2r^2}=-\frac{D'(t)}{(1+a t) D(t)} \equiv C_0\,.
\end{equation}
 Obviously, $C_0$ must be a constant independent of $r$ and $t$. So the only solution is
 \begin{eqnarray}
 c=0\,,
 \end{eqnarray}
 and consequently
 \begin{equation}
 D(t)=D_0\,.
 \end{equation}
Now \eqs{sim4} and \meq{sim5} become
\begin{eqnarray}\label{sim8}
x&=&-2 a^{-1}(a t+1)(D_0-1)\,,\\
y&=&D_0 r\,.\label{sim9}
\end{eqnarray}
Plugging \eqs{sim8} and \meq{sim9} into \eq{sim2}, we have
\begin{equation}
D_0=\frac{2}{3}
\end{equation}
Hence,
\begin{eqnarray}
x&=&\dfrac{2(1+a t)}{3 a}\,,\ \ y=\frac{2}{3} r\,,\\
\xi^a &=&\dfrac{2(1+a t)}{3 a}\ppa{t}{a}+\frac{2 r}{3}\ppa{r}{a}\,.
\end{eqnarray}
Thus,  we have proven that the HMN spacetime is  continuous self-similar only for $c=0$ and $a\neq0$.

\end{CJK*}
\end{document}